\documentclass[11pt]{article} 
\begin{document} 
  \vspace*{1.1cm} 
  \begin{center} 
  {\LARGE \bf Large ``Dipolar" Vacuum Fluctuations \\
  in Quantum Gravity} 
  \end{center} 

  \begin{center} 
  \vskip 10pt 
  Giovanni Modanese\footnote{e-mail address: 
  giovanni.modanese@unibz.it}
  \vskip 5pt
  {\it California Institute for Physics and Astrophysics \\
  366 Cambridge Ave., Palo Alto, CA 94306}
  \vskip 5pt
  and
  \vskip 5pt
  {\it University of Bolzano -- Industrial Engineering \\
  Via Sorrento 20, 39100 Bolzano, Italy}
  \end{center} 

  \baselineskip=.175in 
    
\begin{abstract}
We study a novel set of 
gravitational field configurations, called ``dipolar zero 
modes", which give an exactly null contribution to the
Einstein action and are thus candidates to become large
fluctuations in the quantized theory. They are generated
by static unphysical sources satisfying (up to terms of
order $G^2$) the simple condition $\int d^3x T_{00}
({\bf x})=0$.
We give two explicit examples of virtual sources: 
(i) a ``mass dipole"
consisting of two separated mass distributions with
different signs; (ii) two concentric ``+/- shells". 
The field fluctuations can be large even at
macroscopic scale. There are some, for instance, which
last $\sim 1 \ s$ or more and correspond to the field
generated by a virtual source with size $\sim 1 \ cm$
and mass $\sim 10^6 \ g$. This appears paradoxical, for
several reasons, both theoretical and phenomenological.
We also give an estimate 
of possible suppression effects following the addition 
to the pure Einstein action of cosmological or $R^2$
terms. 

\medskip

\noindent 04.20.-q Classical general relativity.

\noindent 04.60.-m Quantum gravity.

\medskip


\end{abstract}

\section{Introduction}
\label{introd} 
 
Vacuum fluctuations are an essential ingredient of any
quantum field theory, and also in quantum gravity they
play an important role. The presence in the gravitational 
action of a dimensional coupling of the order of
$10^{-33} \ cm$ -- the ``Planck length" -- indicates that
the strongest fluctuations occur at very small scale:
this is the famous ``spacetime foam", first studied by
Wheeler and then by
Hawking and Coleman through functional integral techniques
\cite{haw}.

More recently, Ashtekar and others \cite{ash} analysed the
possible occurrence of large fluctuations in 2+1 gravity
coupled to matter. In this case the theory is classically
solvable and admits a standard Fock-space quantization.
In 3+1 dimensions, however, Einstein quantum gravity is a
notoriously intractable theory, where everything (states, 
transition amplitudes, time...) is highly non-trivial. 

The non-renormalizable UV divergences of the perturbative
expansion may indicate that quantum gravity is not a
fundamental microscopic theory, but an effective
low-energy limit \cite{don}, and will be eventually
replaced by a theory of strings or branes. On the other
hand, it is known from particle physics that the Einstein
lagrangian can be obtained, without any geometrodynamical
assumption, as the only one which correctly accounts
for a gravitational force mediated by helicity-2
particles \cite{alv}. For this reason, it
is important to investigate -- besides the standard
perturbative expansion -- all the basic properties
of the Einstein lagrangian. In the past years we took an
interest into Wilson loops \cite{wil}, vacuum correlations at
geodesic distance \cite{cor}, and the expression of the
static potential through correlations between particles
worldlines \cite{pot}.

In this work we study a set of 
gravitational field configurations, called ``dipolar zero 
modes", which were not considered earlier in the 
literature. They give an exactly null contribution to the
Einstein action, being thus candidates to become large
fluctuations in the quantized theory. We give an explicit 
expression, to leading order in $G$, for some of the field 
configurations of this (actually quite large) set.
We also give an estimate 
of possible suppression effects following the addition 
to the pure Einstein action of cosmological or $R^2$
terms. 

Our zero modes have two peculiar features, which make
them relatively easy to compute: (i) they are 
formally solutions of the
Einstein equations with auxiliary virtual sources;
(ii) their typical length scale is such that they can
be treated in the weak field approximation. We shall see
that these fluctuations can be large even on a
``macroscopic" scale. There are some, for instance, which
last $\sim 1 \ s$ or more and correspond to the field
generated by a virtual source with size $\sim 1 \ cm$
and mass $\sim 10^6 \ g$. This seems paradoxical, for
several reasons, both theoretical and phenomenological.
We have therefore been looking for possible suppression
processes. Our conclusion is that a vacuum energy term
$(\Lambda/8\pi G)\int d^4x \sqrt{g(x)}$ in the action
could do the job, provided it was scale-dependent and
larger, at laboratory scale, than its observed
cosmological value. This is at present only a speculative
hypothesis, however.

The dipolar fluctuations owe their existence to the fact
that the pure Einstein lagrangian $(1/8\pi G) 
\sqrt{g(x)} R(x)$ has indefinite sign also for static
fields. It is well known that the non-positivity of
the Einstein action makes an Euclidean formulation of
quantum gravity difficult; in that context, however,
the ``dangerous" field configurations have small-scale 
variations and could be eliminated, for instance,
by some UV cut-off. This is not the case of the dipolar
zero modes. They exist at any scale and do not make the
Euclidean action unbounded from below, but have instead
null (or $\ll \hbar$) action.

A static virtual source will generate a zero mode provided
it satisfies the condition $\int d^3x T_{00}({\bf x}) =0$
up to terms of order $G^2$. The cancellation of the terms
of order $G$ (Section \ref{s22}) is important from the practical
point of view. In our earlier work on dipolar
fluctuations \cite{pre} we developed some general
remarks based on the form of Einstein equations, and the
result was that in order to generate a zero mode the
positive and negative masses of the source should differ
by a quantity of order $G$, namely $\sim Gm^2/r \sim
mr_{Schw.}/r$; this is very small for weak fields, but
sufficient to produce a ``monopolar" component which
complicates the situation. Explicit calculations in
Feynman gauge now have shown that the terms of order $G$
cancel out exactly. This opens the way
to an amusing ``virtual source engineering" work, to find
explicitly some zero modes and give quantitative 
estimates in specific cases.

When analysing the Wilson loops, we had already pointed 
out some differences in the behavior of gravity and 
ordinary gauge theories, essentially due to the different signs
of the allowed physical sources. Here, again, these
differences are apparent. In gauge theories the real sources
can be both positive and negative; therefore one can close
two Wilson lines at infinity and find the static potential.
The virtual sources cannot give rise to strong static
dipolar fluctuations, because the lagrangian is quadratic
in the fields. On the contrary, in gravity there are no
real negative sources, the potential is always attractive
and Wilson lines cannot be closed; however, since the
lagrangian on-shell is indefinite in sign and equal to
$\sqrt{g(x)}{\rm Tr}T(x)$, we can construct static
zero modes employing +/- virtual sources. Then, of course,
we can Lorentz-boost these modes in all possible ways.

The paper is composed of two main Sections. Section 2
is devoted to the analysis of the dipolar fields and
virtual sources. We start from some general features
and then focus on two examples. Section 3 contains an 
extensive discussion. For a summary of the main contents
see also the {\it Conclusions} Section.

\subsection{Conventions. Sign of $\Lambda$ vs.\ its 
classical effects}
\label{convenz}

Let us define here our conventions. We consider a
gravitational field in the standard metric formalism; the
action includes possibly a cosmological term:
	\begin{eqnarray}
	S & = & S_{Einstein} + S_\Lambda \\
	S_{Einstein} & = & - \frac{1}{16\pi G} \int d^4x
	 \sqrt{g(x)} R(x) \label{e2} \\
	S_\Lambda & = & \frac{\Lambda}{8\pi G} \int d^4x
	 \sqrt{g(x)}
\end{eqnarray}
	with $g_{\mu \nu}(x)=\eta_{\mu \nu}+h_{\mu \nu}(x)$.

By varying this action with respect to $\delta g_{\mu \nu}(x)$ and using
the relation
	\begin{equation}
	\frac{\delta \sqrt{g}}{\delta g_{\mu \nu}} 
	= \frac{1}{2} \sqrt{g} g^{\mu \nu}
\end{equation}
	one finds the field equations
	\begin{equation}
	R_{\mu \nu}(x)- \frac{1}{2} g_{\mu \nu}(x) R(x)  
	+ \Lambda g_{\mu \nu}(x) = -8 \pi G T_{\mu \nu}(x) 
\label{ein1}
\end{equation}

The energy-momentum tensor of a perfect fluid has the form
	\begin{equation}
	[T_{\mu \nu}] = {\rm diag}(\rho,p,p,p)
\end{equation}
	For a zero-pressure ``dust" one has $p=0$.

Now let us introduce a signature for the metric.  Articles in General
Relativity or cosmology use most often the metric with 
signature $(-,+,+,+)$,
and the experimental estimates of $\Lambda$ are mainly referred to this
metric. It is important to fix the sign of the 
cosmological term with reference to the metric signature 
in a way which is clear both formally and intuitively.

If spacetime is nearly flat, we can take
the cosmological term in (\ref{ein1}) to the r.h.s., 
set $g_{\mu \nu}(x)=\eta_{\mu \nu}$ and regard it
as a part of the source. We obtain, in matrix form
	\begin{equation}
	\left[ R_{\mu \nu}- \frac{1}{2} g_{\mu \nu} R 
	\right] = - \left\{ 
	{\rm diag}(-\Lambda,\Lambda,\Lambda,\Lambda) +
	8\pi G {\rm diag}(\rho,p,p,p) \right\} \qquad
	[{\rm metric \ (-,+,+,+)}] 
\label{mm}
\end{equation}
	Which sign
for $\Lambda$ allows to obtain a static solution?  Even without finding
explicitly this solution, we see that for
$\Lambda>0$ the ``pressure" due to the cosmological term is positive and
can sustain the system against gravitational collapse -- especially in the
case of a zero-pressure dust with $p=0$.  At the same time, the
mass-energy density due to the cosmological term is negative and subtracts
from $\rho$, still opposing to the collapse. 

In conclusion, with this convention on the metric signature a static
solution of Einstein equations with a cosmological term can be obtained
for $\Lambda>0$. If we are not interested into a static solution, but into
an expanding space \`a la Friedman-Walker, in that case the effect of a
cosmological term with $\Lambda>0$ will be that of accelerating the
expansion. The most recent measurements of the Hubble constant from Type
Ia supernovae \cite{str,per} suggest indeed that there is a cosmologically
significant positive $\Lambda$ in our universe.

In Quantum Field Theory, on the other hand, the signature 
$(+,-,-,-)$ is more
popular, such that the squared four-interval is $x^2=t^2-|\bf x|^2$. Since
we shall introduce some coupling of gravity to matter fields in the
following, and make a correspondence to the Euclidean case,
we prefer to use this latter convention. We then have, instead
of eq.\ (\ref{mm}) 
	\begin{equation}
	\left[ R_{\mu \nu}- \frac{1}{2} g_{\mu \nu} R 
	\right] = - \left\{ 
	{\rm diag}(\Lambda,-\Lambda,-\Lambda,-\Lambda) +
	8\pi G {\rm diag}(\rho,p,p,p) \right\} \qquad
	[{\rm metric \ (+,-,-,-)}]
\label{mp}
\end{equation}

In this case, a static solution -- or an accelerated expansion --
corresponds to $\Lambda<0$.

\section{The dipolar fluctuations}
\label{fluttu}

We consider the functional integral of pure quantum 
gravity, which represents a sum over all possible field 
configurations weighed with the factor $\exp[i\hbar
S_{Einstein}]$ and possibly with a factor due to the 
integration measure. The Minkowski space is a stationary 
point of the vacuum action and has maximum probability.
``Off-shell" configurations, which are not solutions of 
the vacuum Einstein equations, are admitted in the 
functional integration but are strongly suppressed by 
the oscillations of the exponential factor.

Due to the appearance of the dimensional constant $G$ 
in the Einstein action, the most probable quantum 
fluctuations of the gravitational field ``grow" at very 
short distances, of the order of $L_{Planck}= \sqrt{G
\hbar/c^3} \sim 10^{-33} \ cm$. This led Hawking, 
Coleman and others to depict spacetime at the Planck 
scale as a ``quantum foam" \cite{haw}, with high curvature 
and variable topology. For a simple estimate
(disregarding of course the possibility of topology 
changes, virtual black holes nucleation etc.), suppose 
we start with a 
flat configuration, and then a curvature fluctuation 
appears in a region of size $d$. How much can the 
fluctuation grow before it is suppressed by the
oscillating factor $\exp[iS]$? (We set $\hbar=1$ and
$c=1$ in the following.) A naive dimensional estimate
suggests that $|R|$ should not exceed $\sim G/d^4$,
but in fact only a non-perturbative calculation can
give reliable results in the short-distance regime.
The most accurate estimates of the critical exponents
in lattice quantum gravity are those obtained by Hamber
through the Euclidean Regge calculus \cite{h2000},
and show that the correct behavior in four dimensions is 
	\begin{equation}
	|R| \sim \frac{1}{L_{Planck} d}
\end{equation}
	This is a consequence of the fact that the critical
exponent $\nu$, related to the derivative of the
gravitational $\beta$-function in the vicinity of the
UV fixed point, is very close to 1/3.

\subsection{General features}

There is another way, however, to obtain vacuum field
configurations with action smaller than 1 in natural
units. This is due to the fact that the Einstein action 
has indefinite sign. Consider the Einstein equations 
with a source $T_{\mu \nu}(x)$
	\begin{equation} 
	R_{\mu \nu}(x)- \frac{1}{2} g_{\mu \nu}(x) R(x)  
	= -8 \pi G T_{\mu \nu}(x)
\label{ein} 
\end{equation} 
and their covariant trace 
	\begin{equation} 
	R(x)=8 \pi G {\rm Tr} T(x) =
	8 \pi G g^{\mu \nu}(x) T_{\mu \nu}(x)
\label{tra} 
\end{equation} 
 
Let us consider a solution $g_{\mu \nu}(x)$ of equation 
(\ref{ein}) with a source $T_{\mu \nu}(x)$ obeying the 
additional integral condition 
	\begin{equation} 
	\int d^4x  \sqrt{g(x)}  {\rm Tr}  T(x) = 0
\label{add} 
\end{equation} 
	Taking into account eq.\ (\ref{tra}) we see 
that the pure Einstein action (\ref{e2}) computed for this 
solution is zero. Thus the tensor $T_{\mu \nu}(x)$ only
serves as an auxiliary source in order to construct
zero-modes for the action of pure gravity.
Condition (\ref{add}) can be satisfied by 
energy-momentum tensors that are not identically zero, 
provided they have a balance of negative and positive 
signs, such that their total integral is zero. Of course, 
they do not represent any acceptable physical source, but 
the corresponding solutions of (\ref{ein}) exist 
nonetheless, and are zero modes of the pure Einstein
action.  
 
We shall give two explicit examples of auxiliary sources
(we shall call them ``virtual sources" in the following,
because they generate virtual field configurations): 
(i) a ``mass dipole"
consisting of two separated mass distributions with
different signs; (ii) two concentric ``+/-
shells". In both cases there are some parameters of the
source which can be varied: the total positive and
negative masses $m_\pm$, their distance, the spatial
extension of the sources. 

The procedure for the construction of the zero mode 
corresponding to the dipole is the following. One first 
considers Einstein equations with the virtual source without
fixing the parameters yet. Then one solves them with a 
suitable method, for instance in the weak field 
approximation when appropriate. Finally, knowing 
$g_{\mu \nu}(x)$ one adjusts the parameters in such a 
way that condition (\ref{add}) is satisfied.

\subsection{Computation of $\sqrt{g(x)} g^{00}(x)$}
\label{s22}

Now suppose we have a suitable virtual source, with some free
parameters, and we want to adjust them in such a way
to generate a zero-mode $g_{\mu \nu}(x)$ for which
$S_{Einstein}[g]=0$. We shall always consider static
sources where only the component $T_{00}$ is non
vanishing. The action of their field is
	\begin{equation} 
	S_{zero-mode} = - \frac{1}{2} \int d^4x \sqrt{g(x)}
	g^{00}(x) T_{00}(x)
\end{equation}

To first order in $G$, the
field $h_{\mu \nu}(x)$ generated by a given mass-energy
distribution $T_{\mu \nu}(x)$ is given 
by an integral of the field propagator 
$P_{\mu \nu \rho \sigma}(x,y)$ over the source:
	\begin{equation}
	h_{\mu \nu}(x) = \int d^4y 
	P_{\mu \nu \rho \sigma}(x,y) T^{\rho \sigma}(y)
\label{campo}
\end{equation}
	where in Feynman gauge $P_{\mu \nu \rho 
\sigma}(x,y)$ is given, with our conventions on the 
metric signature, by
	\begin{equation}
	P_{\mu \nu \rho \sigma}(x,y) = \frac{2G}{\pi}   
	\frac{\eta_{\mu \rho}\eta_{\nu \sigma} +
	\eta_{\mu \sigma}\eta_{\nu \rho} -
	\eta_{\mu \nu}\eta_{\rho \sigma}}
	{(x-y)^2 + i\varepsilon}
\end{equation}
	Computing the integral over time in eq.\ 
(\ref{campo}) we obtain for our source
	\begin{eqnarray}
	h_{\mu \nu}({\bf x}) & = & 
	\int_{-\infty}^{+\infty} dy_0 \int d^3y 
	T^{00}({\bf y}) P_{\mu \nu 00}(x,y) \nonumber \\
	& = & \frac{2G}{\pi} 
	(2\eta_{\mu 0} \eta_{\nu 0} - 
	\eta_{\mu \nu} \eta_{00})
	\int_{-\infty}^{+\infty} dy_0 \int d^3y 
	\frac{T^{00}({\bf y})}{(x_0-y_0)^2 -
	({\bf x}-{\bf y})^2 + i\varepsilon} \nonumber \\
	& = & 2G (2\eta_{\mu 0} 
	\eta_{\nu 0} - \eta_{\mu \nu} \eta_{00})
	\int d^3y \frac{T^{00}({\bf y})}{|{\bf x}-{\bf y}|}
\label{acca}
\end{eqnarray}

Thus we have

	\begin{eqnarray}
	\sqrt{g(x)} g^{00}(x) & = & 
	\left[ 1 + \frac{1}{2} {\rm Tr}h({\bf x}) 
	+ o(G^2) \right] 
	\left[ 1 + h^{00}({\bf x}) + o(G^2) \right] 
	\nonumber \\
	& = & 1 + \frac{1}{2} {\rm Tr}h({\bf x}) 
	+ h^{00}({\bf x}) + o(G^2) 
	\nonumber \\
	& = & 1 + 2G \left[ \frac{1}{2}
	(2\eta_{\mu 0} \eta_{\nu 0} - 
	\eta_{\mu \nu} \eta_{00})\eta^{\mu \nu}
	+ (\eta_{00})^2 \right]
	\int d^3y \frac{T^{00}({\bf y})}{|{\bf x}-{\bf y}|}
	+ o(G^2)
	\nonumber \\
	& = & 1 + o(G^2)
\label{cancel}
\end{eqnarray}
	and finally the action is
	\begin{equation} 
	S_{zero-mode} = - \frac{1}{2}
	\int d^4x T_{00}(x) + o(G^2)
\label{e17}
\end{equation}
	
Therefore provided the integral of the mass-energy density
vanishes, the action of our field configuration is of
order $G^2$, i.e., practically negligible, as we 
check now with a numerical example. Let us choose 
the typical parameters of the source as follows:
	\begin{eqnarray}
	r & \sim & 1 \ cm \nonumber \\
	m & \sim & 10^k \ g \simeq 10^{37+k} \ cm^{-1}
\label{esempio}
\end{eqnarray}
	(implying $r_{Schw.}/r \sim 10^{-29+k}$). 
We assume in general an adiabatic switch-on/off of the
source, thus the time integral contributes to the action
a factor $\tau$. We shall keep $\tau$ (in natural units)
very large, in order to preserve the static character 
of the field. Here, for instance, let us take
$\tau \sim 1\ s \simeq 3 \cdot 10^{10} \ cm$.
With these parameters we have
	\begin{equation}
	S_{zero-mode}^{order \ G^2} 
	\sim \tau \frac{G^2 m_\pm^2}{r^3}
	\sim 10^{-20+3k}
\end{equation}  
	Thus the field generated by a virtual source 
with typical size (\ref{esempio}), satisfying the condition
$\int d^3x T^{00}({\bf x}) = 0$, has negligible action
even with $k=6$ (corresponding to apparent matter 
fluctuations with a density of $10^6 \ g/cm^3$ !) 
This should be compared to the huge action of the field
of a {\it single}, unbalanced virtual mass $m$; with 
the same values we have
	\begin{equation}
	S_{single \ m} = - \frac{1}{16\pi G} \int d^4x 
	\sqrt{g(x)} R(x) = - \frac{1}{2} 
	\int d^4x \sqrt{g(x)} 
	{\rm Tr} T(x) \sim \frac{1}{2} \tau m + o(G^2)
	\sim 10^{47+k}
\end{equation}

This example shows that the cancellation of the first 
order term in (\ref{cancel}) allows to obtain a simple
lower bound on the strength of the fluctuations.
In principle, however, one could always find all the terms
in the classical weak field expansion, proportional to
$G$, $G^2$, $G^3$, etc., and adjust $T_{00}$ as to have 
$S_{zero-mode} =0$ exactly. They can be represented by
those Feynman diagrams of perturbative quantum gravity
which contain vertices with 3, 4 ... gravitons but do not 
contain any loops. 
The ratio between each contribution to $S$ and that of
lower order in $G$ has typical magnitude
$r_{Schw.}/r$,
where $r_{Schw.}=2\pi G m_\pm$ is the Schwarzschild radius
corresponding to one of the two masses and $r$       
is the typical size of the source. For a wide range of
parameters, this ratio is very small, so the expansion
converges quickly. From now on we agree that the  
``$o(G^2)$" term in eq.\ (\ref{e17}) comprises all the terms
quadratic in the field, like for instance that arising
from the expansion of $\sqrt{g(x)}$.

\subsection{Explicit examples of static virtual sources}

\noindent
{\it (i) The mass dipole}

As an example of unphysical source which satisfies 
(\ref{add}) one can consider the static field produced 
by a ``mass dipole". Certainly negative masses do not 
exist in nature; here we are interested just in the formal 
solution of (\ref{ein}) with a suitable $T_{\mu \nu}$, 
because for this solution we have $\int d^4x  \sqrt{g} 
R=0$.  Let us take the following $T_{\mu \nu}$ of a 
static dipole centered at the origin ($m_{+},m_{-}>0$): 
	\begin{equation} 
	T_{\mu \nu}({\bf x}) = \delta_{\mu 0}  
	\delta_{\nu 0}  
	\left[ \frac{m_{+}}{r_{+}^3} f_{+}({\bf x})  
	- \frac{m_{-}}{r_{-}^3} f_{-}({\bf x}) \right] 
\label{dip} 
\end{equation} 
	where 
	\begin{equation} 
	f_{\pm}({\bf x}) \equiv f \left( 
	\frac{{\bf x}\pm{\bf a}}{r_\pm} \right)
\end{equation} 
	and $f({\bf x})$ is a smooth test function with 
range $\sim 1$ and normalized to 1, which represents the 
mass density. Thus we have a positive source of mass 
$m_{+}$ and radius $r_{+}$ (placed at ${\bf x}=-{\bf a}$) 
and a negative source with mass $-m_{-}$ and radius 
$r_{-}$ (placed at ${\bf x}={\bf a}$). The radii of the 
two sources are such that $a \gg r_\pm \gg r_{Schw.}$, 
where $r_{Schw.}$ is the Schwarzschild radius corresponding 
to the mass $m_{+}$. 
 
The mass $m_{-}$ is in general slightly different from 
$m_{+}$ and chosen in such a way to compensate the
small difference, due to the $\sqrt{g}g^{00}$ factor, 
between the integrals 
	\begin{equation} 
	I_+ = \int d^4x  \sqrt{g(x)} g^{00}(x)
	\frac{f_+({\bf x})}{r_+^3} \qquad {\rm and} 
	\qquad
	I_- = \int d^4x \sqrt{g(x)} g^{00}(x) 
	\frac{f_-({\bf x})}{r_-^3}
\label{piu} 
\end{equation} 
 
The action of the dipole is
	\begin{equation} 
	S_{Dipole} = - \frac{1}{2} \int d^4x 
	T_{00}({\bf x}) =
	-\frac{1}{2} \tau (m_+ - m_-) + o(G^2)
\end{equation} 
	The condition for $S_{Dipole}=0$ is $m_+ = m_-$,
apart from terms of order $G^2$ (i.e., our dipoles have 
in reality a tiny monopolar component).

Also note that the values of the masses and the radii
$r_\pm$ (both of order $r$) can vary in a
continuous way -- under the only condition that
$m_+ = m_-$. This implies that these
(non singular) ``dipolar" fields constitute a 
subset with nonzero volume in the 
functional integration. Actually, they are only a subset
of the larger class of solutions of the Einstein equations
with sources satisfying eq.\ (\ref{add}).

\medskip
\noindent
{\it (ii) The concentric +/- shells}

Consider two concentric spherical shells in contact, the internal
one with radii $r_1$, $r_2$, and the external one with
radii $r_2$, $r_3$ ($r_1 < r_2 < r_3$). Let the internal
shell have mass density $\rho_1$ and the external shell
density $\rho_2$, with opposite sign. 
The condition for zero action requires,
up to terms of order $G^2$, that the total positive mass
equals the total negative mass, i.e.,
	\begin{equation} 
	\rho_1 (r_2^3 - r_1^3) + \rho_2 (r_3^3 - r_2^3)
	= 0
\label{e25}
\end{equation} 
	(more generally, if the densities $\rho_1$ and
$\rho_2$ are not constant throughout the shells, one has
a suitable integral condition). 

The spherical symmetry of the corresponding field
configuration offers some advantages when one computes
the contributions to the cosmological term and the
Newtonian self energy (compare Sect.\ \ref{s31}).

\subsection{Contribution of virtual dipoles
to the cosmological and $R^2$ terms} 
\label{stima}
 
In the previous Sections we have seen that the pure Einstein
action admits zero-modes having the form of virtual
dipole field configurations with a small monopole
residual. These field configurations are characterized
by the parameters $r_\pm$ (radii of the virtual +/-
sources), $a$ (distance between the sources) and $m_\pm$
(masses of the sources). We worked out these
configurations as solutions of the
linearized Einstein equations. We also checked that the
weak field approximation is appropriate in a whole 
``macroscopic" range of the parameters $r_\pm$, $a$ and $m$.
This is possible because these configurations (unlike the 
spacetime foam at the Planck scale) yield $\int d^4x
\sqrt{g(x)}R(x) = 0$ thanks to a cancellation between
the $R$ contributions in two distinct regions of space.
Similar considerations can be done for the field of the
concentric +/- shells.

It is natural to ask whether the dipolar fluctuations
can be suppressed by other terms present in the 
gravitational action besides the pure Einstein term.
Possible candidates are the $R^2$ terms (usually relevant,
however, only at very small distance) and the cosmological
term. Let us first look at the latter (see also
our general remarks on the role of a cosmological constant in
quantum gravity in Section \ref{idea}).

When a static source is spherically symmetric, we
can use {\it outside} it the exact Schwarzschild metric with
invariant interval
	\begin{equation}
	ds^2 = \left( 1 - \frac{2GM}{r} \right)^{-1} dt^2
	- \left[ \left( 1 - \frac{2GM}{r} \right) dr^2 +
	r^2 d\theta^2 + r^2 \sin^2 \theta d\phi^2 \right]
\end{equation}
	The determinant of this metric equals that of flat
space, so the presence of one single 
spherically-symmetric source does not change
the volume of the outer space and does not contribute
to the cosmological term.

We shall therefore handle separately the cases of the
mass dipole and the concentric +/- shells.

\medskip
\noindent
{\it The mass dipole}

In the linearized approximation the integral
$S_\Lambda = (\Lambda/8\pi G)\int d^4x \sqrt{g(x)}$
for a dipolar fluctuation can be splitted into the sum
of the integrals of the field $h_{+}({\bf x})$ generated
by the positive mass and the field $h_{-}({\bf x})$ 
generated by the negative mass. Both fields are
spherically symmetric, thus there is no contribution
of order $G$ to $S_\Lambda$ outside the sources.

To order $h^2 \sim (Gm)^2$ the field outside the sources
differs from the sum of their Schwarzschild fields, and
we do have some contributions to the cosmological term,
but they are very small. One finds, inserting the numerical
values (\ref{esempio}) and the current estimate for
$|\Lambda| G$, namely $|\Lambda| G \sim 10^{-116}$
	\begin{equation}
	\Delta S_{\Lambda,outside} \sim \tau^2 |\Lambda| G m^2
	\sim 10^{-22+2k}
\end{equation}

On the other hand, the integrals of $\sqrt{g(x)}$ {\it
inside} the sources contribute to the action already at first
order in $h_{\mu \nu}$. Let us use the explicit solutions
in Feynman gauge found in the previous section and
disregard the effect of the positive source inside
the negative one and viceversa. 
(This will give small corrections proportional to
$a/r_\pm$, but does not change the magnitude orders.)
We denote by $\omega({\bf x})$ the
characteristic function of a 3-sphere with unit radius placed
at the origin of the coordinates, and define
	\begin{equation}
	\omega_\pm({\bf x}) \equiv \omega \left( 
	\frac{{\bf x}\pm{\bf a}}{r_\pm} \right)
\end{equation}
	We then have, to leading order
	\begin{eqnarray}
	\Delta S_{\Lambda,inside} & = & 
	\frac{\Lambda}{8\pi G} 
	\left[ \int d^4x \frac{1}{2}
	{\rm Tr}h_{+}({\bf x}) \omega_+({\bf x})+
	\int d^4x \frac{1}{2}
	{\rm Tr}h_{-}({\bf x}) \omega_-({\bf x})\right]
	\nonumber \\                
	& = & \frac{\Lambda}{8\pi G} \frac{\tau}{2}
	\int d^3x \omega_+({\bf x}) \left( 
	\frac{-4m_{+}G}{r_+^3} \right)
	\int d^3y \frac{f_+({\bf y})}{|{\bf x}-{\bf y}|} +
	\nonumber \\                
	& & +\frac{\Lambda}{8\pi G} \frac{\tau}{2}
	\int d^3x \omega_+({\bf x}) \left( 
	\frac{4m_{-}G}{r_-^3} \right)
	\int d^3y \frac{f_-({\bf y})}{|{\bf x}-{\bf y}|}
\end{eqnarray}
In the double integrals we can suitably shift the variables
by $\pm {\bf a}$ and re-scale them as 
${\bf x} \to {\bf x}'r_\pm$, ${\bf y} \to {\bf y}'r_\pm$,
obtaining a pure number $\xi$ of order 1 multiplied by 
$r_+^5$ and $r_-^5$, respectively. Finally we obtain
	\begin{equation}
	\Delta S_{\Lambda,inside} = -\frac{\xi}{4\pi}
	\tau \Lambda (m_{+}r_+^2 - m_{-}r_-^2)
\label{e30}
\end{equation}
	with
	\begin{equation}
	\xi = \int d^3x' \int d^3y' 
	\frac{\omega({\bf x}') f({\bf y}')}
	{|{\bf x}'-{\bf y}'|}
\end{equation}

With the usual values
we find, apart from an adimensional constant of order 1
	\begin{equation}
	\Delta S_{\Lambda,inside} \sim
	10^{-3+k}
\end{equation}
	This means that a relatively small increase 
in the value of
$|\Lambda|$ would be sufficient to suppress the strongest
fluctuations (except for those with $r_+=r_-$ exactly).

\medskip
\noindent
{\it The concentric +/- shells}

In this case $\Delta S_{\Lambda,outside}$ vanishes exactly.
Inside the source we have to leading order
	\begin{equation}
	\Delta S_{\Lambda,inside} = 
	\frac{\Lambda \tau}{8\pi G} \frac{1}{2}
	\int d^3x {\rm Tr} h({\bf x})
\end{equation}

Since ${\rm Tr} h({\bf x}) = 4 V_{Newt.}({\bf x})$
(compare eq.\ (\ref{acca})), the integral is a special case
of one we shall compute in Section \ref{s31} The result is
	\begin{equation}
	\Delta S_{\Lambda,inside} = 
	\Lambda \tau m r^2 Q(\beta)
\end{equation}
	where $r_2 \equiv r$, $r_3 \equiv \beta r$ and 
$Q(\beta)$ is an adimensional polynomial which can be either
positive or negative, depending on the ratio 
$|\rho_1|/|\rho_2|$. The magnitude order is the
same as for the dipole.

\medskip

Finally, a word about the $R^2$ term. It is typically
of the form
	\begin{equation}
	S_{R^2} = \alpha \int d^4x \sqrt{g(x)} R^2(x)
\end{equation}
	where $\alpha$ is a (small) adimensional coupling and
$R^2$ can be replaced by more complex scalars like
$R_{\mu \nu \rho \sigma}R^{\mu \nu \rho \sigma}$ etc.
For an order of magnitude estimate it suffices to multiply
the square of the curvature in the sources, namely
$R^2 \sim (G{\rm Tr}T)^2 = (Gm/r^3)^2$ by their volume
$V^{(4)} \sim \tau r^3$. We find in this way, still with
the same parameters,
	\begin{equation}
	S_{R^2} \sim \alpha \tau G^2 \frac{m^2}{r^3} \sim
	\alpha 10^{-48+2k}
\end{equation}
Thus the allowed values for $m$ are very large, i.e., there
is no significant suppression of the virtual dipoles by 
the $R^2$ terms at this scale.

\section{Discussion}

\subsection{Why are these fluctuations paradoxical}
\label{s31}

The order of magnitude estimates given in the previous
Section show that the dipolar vacuum
fluctuations allowed in the functional integral
formulation of pure Einstein quantum gravity 
(i.e., such to give $S \ll 1$ in natural units) are
very intense also at macroscopic scale. 

One may think that such large fluctuations, if real, 
would not remain unnoticed. Even though vacuum fluctuations
are homogeneous, isotropic and Lorentz-invariant, they
could manifest themselves as noise of some kind. 
Most authors are skeptic about the possibility
of detecting the noise due to spacetime foam \cite{ame,ell},
but the virtual dipole fluctuations described in this
paper are much closer to the laboratory scale.
Observable quantities, like for instance the invariant
intervals $ds^2=g_{\mu \nu}dx^\mu dx^\nu$ and the
connection coefficients $\Gamma^\rho_{\mu \nu}$
could then exhibit strong fluctuations.

The existence of these
fluctuations would be paradoxical, however, already 
at the purely conceptual level. Common wisdom in particle
physics states that the vacuum fluctuations in free space
correspond to virtual particles or intermediate states 
which live very short, i.e., whose
lifetime is close to the {\it minimum} allowed by the Heisenberg
indetermination relation. Let us first give a brief
formal justification of this rule, and then
compare it to our dipole fluctuations. 

It is often the case that a quantum field theory has
an imaginary time formulation, where the (positive-definite)
lagrangian density corresponds to the original hamiltonian 
density $H$. For a scalar field, for instance, one has
$H=(1/2)[(\partial \phi)^2 + ({\rm grad} \phi)^2
+ m^2 \phi^2]$ and the Euclidean functional integral is
given by $z_{Eucl} = \int d[\phi] \exp[-\int dt \int d^3x
H(t,{\bf x})]$. A field fluctuation localized to a region
of size $\tau V^{(3)}$ is weighed in the functional
integral by the factor $\exp[-\tau V^{(3)}H]=\exp[-\tau E]$
and is thus effectively suppressed unless approx.\ 
$\tau E < 1$. Another notable example is the electromagnetic
field. Also in this case the analytical continuation of the lagrangian
$L = (-1/8\pi)[{\bf E}^2-{\bf B}^2]$ yields the energy
density $H = (1/8\pi)[{\bf E}^2+{\bf B}^2]$; to check
this, one just needs to impose the $A_0=0$ gauge and
remember that only the electric field contains time 
derivatives of ${\bf A}$. 

Now let us estimate the product
$E\tau$ for the dipolar fluctuations. The total energy of
a static gravitational field configuration vanishing at
infinity is the ADM energy. Since the source of a
dipolar fluctuation satisfies the condition $\int d^3x
T_{00}({\bf x})=0$ up to terms of order $G^2$, the
dominant contribution in the ADM energy is the Newtonian
binding energy \cite{mur}.

The binding energy of the 
field generated by a source of mass $m$ and size
$r$ is of the order of $E \sim -Gm^2/r$, where
the exact proportionality factor depends on the details
of the mass distribution. 
For a dipolar field configuration characterized by
masses $m_+$ and $m_-$ 
and radii of the sources $r_+$ and $r_-$,
the total gravitational energy is of the order of
	\begin{equation}
	E_{tot} \sim - G m_\pm^2 \left( \frac{1}{r_-}
	+ \frac{1}{r_+} \right)
\end{equation}
	(disregarding the interaction energy between the two
sources, proportional to $1/a \ll 1/r$).
For an order of magnitude estimate with the
parameters (\ref{esempio}) we can suppose that 
$r_+$ and $r_-$ are both of the order of
$1 \ cm$. We then have $E_{tot} \sim G m_\pm^2 \sim
10^{12+k} \ cm^{-1}$. Remembering that 
$k$ can take values up to $k=6$,
we find for these dipolar fluctuations $\tau E_{tot} \sim
10^{28}$! 

(For comparison, remember the case of a
``monopole" fluctuation of virtual mass $m$ and duration $\tau$. The 
condition $S<1$ implies $\tau m<1$. The dominant contribution
to the ADM energy is just $m$, thus the rule $E\tau<1$ is
respected.)

The Newtonian binding energy of the concentric +/- shells
is given, like in electrostatics, by the formula
$E = (1/2) \int d^3x \rho({\bf x})V_{Newt.}({\bf x})$.
For general values $r_1$, $r_2$, $r_3$ of the radii
and $\rho_1$, $\rho_2$ of the densities (constrained
by the zero total mass condition (\ref{e25})), one obtains
a complicated expression, namely
	\begin{eqnarray}
	E & = & \frac{\pi \rho_1}{90r_2^2} \{
	\rho_2 (r_2^2 + r_2 r_3 + r_3^2) (6r_2^5
	- 15 r_2^4 r_3 + 10 r_2^3 r_3^2 - r_3^5)
	- \rho_1 [9r_1^7 - 11 r_1^6 r_2 - r_1^5 r_2^2    
	- 10 r_1^4 r_2^3 + \nonumber \\ 
	& & + 5 r_1^3 ( r_2^4+2 r_2^3 r_3 + 2 r_2 r_3^3  
	- 2 r_3^4) + r_1^2 r_2^5 + r_1 r_2^6 +
	2 r_2^3 (3 r_2^4 - 5 r_2^3 r_3 - 5 r_2 r_3^3
	+ 5 r_3^4)]\}
\end{eqnarray}
	We can study the sign and magnitude of $E$
setting $r_2=r$, $r_1=\alpha r$ ($0<\alpha<1$) and
$r_3=\beta r$ ($\beta>1$). We express $\alpha$
in terms of $\beta$ using (\ref{e25}) and finally obtain
	\begin{equation}
	E = \frac{Gm}{r} P(\beta)
\end{equation}
	where $P(\beta)$ is a polynomial which is positive
if $|\rho_1|>|\rho_2|$ (the repulsion between the two shells
predominates) and negative if $|\rho_1|<|\rho_2|$ (the 
attraction inside each shell predominates).

This result is quite interesting, because

(i) Unlike the formula for the energy of the dipolar field,
it does not contain any approximation to order $G$.

(ii) From the physical point of view it is reasonable
to admit -- remembering that we are in a weak-field regime
and forgetting general covariance for a minute -- that
the binding energy is localized within the surface of
the outer shell (the field is $o(G^2)$ outside). The
energy density is therefore of the order of
$\frac{|E|}{r^3} \sim \frac{Gm}{r^4} \sim 10^{29+k} \ 
cm^{-4}$ (with the parameters (\ref{esempio})), and can take both
signs. This value looks quite large, even though the
Ford-Roman inequalities \cite{for} or similar bounds do
not apply to quantum gravity, where the metric is not 
fixed but free to fluctuate, and there is in general no 
way to define a local energy density (except outside the
sources -- see \cite{mash}).

\subsection{A scale-dependent $\Lambda$ ?}
\label{idea}

We have seen that a vacuum energy or cosmological term 
in the gravitational action is able to cut-off part of
the dipolar fluctuations. This works better at large
scales, because the $\Lambda$-term does not contain any field
derivatives. We may also hypothesize that the effective
value of $\Lambda$ at scales of the order of $1 \ cm$
is larger than the value observed at cosmological scale.
In the following we summarize some theoretical
arguments supporting this idea. One would have, in other
words, a small, negative, scale-dependent $\Lambda_{eff}$,
a sort of residual of purely gravitational
self-adjustment processes taking place at the Planck
scale.

We already mentioned the role played by the cosmological 
constant at the classical level. In particular, looking
for solutions of Einstein equations of 
the Friedman-Robertson-Walker
type, i.e. with an expanding space, one finds well-defined relations
between the Hubble constant, the density of various kinds of matter, and
$\Lambda$ \cite{str,per}. In the last years, most
estimates have given a {\it negative} value $\Lambda$ 
(in our conventions) of the order of $10^{-50} \ cm^{-2}$.

The effect of a cosmological term in the quantum field theory of gravity
is less clear. On one hand, there are some ``naive" expectations; on the
other hand, formal results which are 
however difficult to interpretate. 

The naive view consists in disregarding the effect of the cosmological
term on the global geometry of spacetime, as compared to the effect of
matter or pre-existing (null) curvature. Therefore one just expands the
gravitational action around a flat background and studies quantum
fluctuations. These are determined to leading order by the part of the
action quadratic in $h_{\mu \nu}$. In spite of the different tensorial
form of the Einstein term $\int dx \sqrt{g}R$ and the cosmological term
$\int dx \sqrt{g}$, their quadratic parts are similar. In Feynman gauge
they are both proportional to the quantity $\left[ 2 {\rm Tr}h^2 - ({\rm
Tr}h)^2 \right]$, multiplied by 
$\partial^\mu \partial_\mu$ in the case of the Einstein
term and by $\Lambda$ in the case of the cosmological term. Thus in this
approximation the cosmological term corresponds to a mass term for the
graviton; the mass is real for $\Lambda < 0$ and 
imaginary for $\Lambda >
0$ (in our conventions -- see Section \ref{convenz}). This implies respectively a finite range
propagator, \'a la Yukawa, with range of the order of 
$|\Lambda|^{-1/2}$, or
the existence of unstable modes growing in time like 
real exponentials \cite{vel}.
Intuitively, the reason for this behavior is clear (see also
\cite{gro}), because a positive $\Lambda$ corresponds to a positive
mass-energy density, which is gravitationally unstable.

In {\it pure quantum gravity} the curvature of the classical background is solely
determined by $\Lambda$, and therefore the previous 
approach does not really
make sense. For instance, if $\Lambda < 0$, then the solution of the
classical Einstein equations is a spacetime with curvature radius of the
order of $\Lambda^{-1/2}$; the Yukawa range predicted by the flat space
expansion would then coincide with the size of the universe. There have
thus been some attempts at quantizing the gravitational action with
respect to a background with constant curvature (de Sitter or anti-de
Sitter). The theory is mathematically very difficult 
\cite{woo}; there is some
evidence, however, that the graviton stays massless, while novel strong
infrared effects would arise (due to the dimensional self-coupling
$\Lambda$), which might force the renormalized value of $\Lambda$ to
``relax towards zero". 

The {\it Euclidean theory of pure quantum gravity} 
is obtained from the Lorentzian theory in our conventions with
the standard analytical continuation $t_{Lor} 
\to -it_{Eucl}$. In the lattice approach in 4D
\cite{hwh}, $G$ and $\Lambda$ are entered as bare couplings
at the beginning, and then the discretized space evolves
according to a Montecarlo algorithm. Unlike in
perturbation theory, where a flat background is introduced
by hand, here flat space appears dynamically; namely, the
average value of the curvature is found to vanish on a
transition line in the bare-couplings space. This line
separates a ``smooth-phase", with small negative curvature,
from a ``rough", collapsed, unphysical phase, with large
positive curvature. The collapse can be understood
observing that the cosmological action is of the form
$\Lambda V^{(4)}$, where $V^{(4)}$ is the volume of the 
lattice,
thus when $\Lambda_{eff}=\langle R \rangle$ is positive,
the volume tends to decrease.

It turns out that as the continuum limit is approached, the
adimensional product $|\Lambda_{eff}|G_{eff}$ behaves like
	\begin{equation}
|\Lambda_{eff}|G_{eff} \sim (l_0/l)^\gamma
\label{e7}
\end{equation}
	where $l$ is the scale, $l_0$ is the lattice spacing, $\gamma$ a
critical exponent and the sign of $\Lambda_{eff}$ is 
negative. Furthermore, one can reasonably assume
that $l_0 \sim L_{Planck}$, and that the
scale dependence of $G_{eff}$ is much weaker than that of $\Lambda_{eff}$.

A scale dependence of $\Lambda_{eff}$ like that in eq.\ (\ref{e7}) also
implies that any bare value of $\Lambda$, expressing the energy density
associated to the vacuum fluctuations of the quantum fields including the
gravitational field itself, approaches zero at long distances just by
virtue of the gravitational dynamics, without any need of a fine tuning.
One would have, in other words, a purely gravitational solution of the
cosmological constant problem.

It is remarkable that the conclusions of Euclidean
lattice theory concerning the instability with $\Lambda > 0$ agree
qualitatively with those obtained in the naive approach;  and this in
spite of the fact that the above argument concernig the volume of
spacetime does not hold in the Lorentzian theory 
because in this case both positive and negative volume variations
are suppressed by the oscillating factor $\exp[iS]$ in the
functional integral.

\subsection{Local changes in $\Lambda$}
\label{coerente}

The ability of the $\Lambda$-term to cut-off part of 
the dipole fluctuations has an inevitable consequence.
Consider the coupling of gravity to a scalar field
$\phi$, with lagrangian density
	\begin{equation}
	L = \frac{1}{2} \left( \partial_\alpha \phi 
	\partial^\alpha \phi - m^2 \phi^2 \right) =
	\frac{1}{2} \left[ \left( 
	\frac{\partial \phi}{\partial t} \right)^2
	- ({\rm grad} \phi)^2 - m^2 \phi^2 \right]
\end{equation}
and energy-momentum tensor
	\begin{equation}
	T_{\mu \nu} = \Pi_\mu \phi 
	\partial_\nu \phi - g_{\mu \nu} L =
	\partial_\mu \partial_\nu \phi - g_{\mu \nu} L
\end{equation}

The interaction term in the gravitational action is        
	\begin{equation}
	S_{matter} = \frac{1}{2} \int d^4x
	 \sqrt{g(x)} T^{\mu \nu}(x) h_{\mu \nu}(x)
\end{equation}
	and to lowest order in $h_{\mu \nu}$ we have
	\begin{equation}
	S_{matter} = \frac{1}{2} \int d^4x \left(
	h_{\mu \nu} \partial^\mu \phi \partial^\nu \phi
	- {\rm Tr}h  L \right)
\end{equation}

On the other hand, the cosmological action is, still 
to lowest order in $h_{\mu \nu}$ and
expanding $\sqrt{g} = 1 + \frac{1}{2} {\rm Tr}h + ...$
	\begin{equation}
	S_\Lambda = \frac{\Lambda}{8\pi G} \int d^4x
	\left( 1 + \frac{1}{2} {\rm Tr}h \right)
\end{equation}

We can say that to lowest order the 
coupling of gravity to the field $\phi$
produces a typical source term for $h_{\mu \nu}$, 
of the form $h_{\mu \nu}
\partial^\mu \phi \partial^\nu \phi$, and subtracts 
from the cosmological
constant $\Lambda$ the local density $8\pi GL(x)$, 
because we can write, apart from an additive constant,
	\begin{equation}
	S_{matter} + S_\Lambda = \frac{1}{2} \int d^4x
	h_{\mu \nu} \partial^\mu \phi \partial^\nu \phi
	+ \frac{1}{2} \int d^4x  {\rm Tr}h
	\left( \frac{\Lambda}{8\pi G} - L \right)
\end{equation}

This separation of the matter coupling in two parts looks
in general quite arbitrary, but it can be useful if the
lagrangian density is such to affect locally the
``natural" cosmological term and set free gravitational
fluctuations corresponding to virtual mass densities
much larger than the real density of the field $\phi$.

An example will clarify our point. Suppose that $\phi$
represents some coherent fluid with the density of
ordinary matter ($\sim 1 \ g/cm^3$). We have seen that at
the scale of 1 $cm$ the dipolar fluctuations are cut-off
according to eq.\ (\ref{e30}). For $\Lambda$ equal to the
cosmologically observed value of $\sim 10^{-50} \ cm^{-2}$,
the exponent $k$ can take values up to $k=3$, 
corresponding to fluctuations with virtual sources of
density $\sim 10^3 \ g/cm^3$. (This is a prudent
estimate; for short-lasting fluctuations -- less than 
$1\ s$ -- and for those with $r_-=r_+$, the virtual mass
density can be even higher.)

If the value of $L$ in some region is comparable to
$\Lambda/8\pi G$, this can introduce
an inhomogeneity in the cut-off mechanism. The result will
be a local inhomogeneity of the dipolar fluctuations, which,
given their strength, could dominate the effects of the
coupling $h_{\mu \nu} \partial_\mu \phi 
\partial_\nu \phi$ to the real matter.

Note that the magnitude of $L$ depends 
on whether $\phi$ is itself
``on shell" or not. For a free, spatially homogeneous 
scalar field, for instance, the Klein-Gordon equation implies 
$\phi = const. \cdot e^{\pm imt}$. Therefore {\it on shell} one
has $L=0$ exactly, even though the single terms in 
$L$ can well be (for atomic-scale masses and gradients) 
of the order of $10^{33} \ cm^{-4}$.

The mechanism sketched above also has an Euclidean analogue
\cite{rol}, but a better understanding of the dipolar 
fluctuations is necessary before any progress in this
direction can be made.

\section{Conclusions}

In the first part of this work we have studied the general
features of ``dipolar" zero modes of the pure Einstein
action, giving some explicit examples in the weak-field
approximation. We used a method based upon the classical
Einstein equations with suitable virtual sources. Our aim
was to prove in a rigorous way the null-action property
of these modes. For applications to the quantum case
we made reference to the (Lorentzian) functional integral.
This represents just one of the possible approaches to
quantum gravity, but in fact also the Planck-scale
fluctuations have been studied through integral functional
techniques \cite{haw,h2000,hwh}. It should be stressed that the
numerical estimates presented in Section \ref{s22} are only
lower limits based on specific examples. The strength
of the fluctuations can be in general larger.

In the Discussion Section we have been less concerned with
rigor. We have described some paradoxical features of the large
dipole fluctuations, and possible suppression processes.
The ADM energy of the dipolar fields can be both positive
and negative, and turns out to be very large compared to
their inverse duration
$\tau^{-1}$. If we admit (as is quite reasonable for
the +/- shells) that this energy is localized, the
corresponding density appears large, too -- even though the
Ford-Roman inequalities or similar bounds do not apply 
to quantum gravity, where the metric is not fixed. 

The hypothesis of a scale-dependent cosmological constant
remains at present speculative, yet only the
$\Lambda$-term seems to be capable of suppression at
large scales. From the purely phenomenological point of
view, the existence of a negative (in our conventions)
$\Lambda_{eff}$, which reduces to the observed
$\Lambda \sim 10^{-50} \ cm^{-2}$ at cosmological scale
but is some orders of magnitude larger at $cm$ scale,
is probably less disturbing than the existence of
large quantum fluctuations \cite{pio}.

Independently from the effective-$\Lambda$ hypothesis,
the results of Sections \ref{stima} and \ref{coerente} show that any local
vacuum term of the form $g_{\mu \nu}(x)L(x)$ acts as
a cutoff for the dipolar fluctuations, especially for
those at large scale. This can cause local inhomogeneities,
which are usually important when dealing with vacuum
fluctuations in quantum field theory, and deserve
further investigation.

\bigskip
{\bf Acknowledgments} - This work was supported in part 
by the California Institute for Physics
and Astrophysics via grant CIPA-MG7099.
The author is grateful to C. Van Den Broeck for useful 
discussions.

\end{document}